# A method to test HF ray tracing algorithm in the ionosphere by means of the virtual time delay

*Cesidio Bianchi[a], Alessandro Settimi[a], Carlo Scotto[a], Adriano Azzarone[a], Angelo Lozito[a].*

[a] Istituto Nazionale di Geofisica e Vulcanologia, V. di Vigna Murata, 605 - 00143 Rome, Italy,

*Corresponding author:*

cesidio.bianchi@ingv.it

## Abstract

It is well known that a 3D ray tracing algorithm furnishes the ray's coordinates, the three components of the wave vector and the calculated group time delay of the wave along the path. The latter quantity can be compared with the measured group time delay to check the performance of the algorithm. Simulating a perfect reflector at an altitude equal to the virtual height of reflection, the virtual time delay is assumed as a real time delay. For a monotonic electronic density profile we find a very small relative difference between the calculated and the virtual delay for both analytic and numerical 3D electronic density models.

# 1. Introduction

According to the ray theory, Three-Dimensional (3D) ray tracing algorithms calculate the coordinates reached by the wave vector and its three components. Another important quantity is the calculated group time delay $t_{calc}$ of the wave along the path. The group time delay is a useful quantity that allows checking of the performance of a ray tracing algorithm when a 3D ionospheric model is assumed correct. In case of a real measurement, as in Over the Horizon (OTH) radar applications or in backscattering ionospheric soundings (or perfectly synchronized oblique sounding experiment), it is possible to compare the calculated time delay $t_{calc}$ with the measured time delay $t_{real}$. When the differences between the calculated and the measured time delay are within an acceptable error mainly due to discrete step of the mathematical process and the numerical 3D ionospheric model, we assume that both 3D ray tracing algorithm and 3D ionospheric model work properly. Nevertheless, the measured delay is usually not available for ionospheric 3D ray tracing users since real measurements of this quantity are quite uncommon. In order to test the algorithm performance, an alternative method to obtain something similar to the term $t_{real}$ is thus required. Propagation theorems and refraction laws, widely used in ionospheric physics applications, have been exploited to reach this goal. Under the assumption of a flat layered ionosphere we imagine to put a perfect reflector at an altitude equal to the vertical virtual height of reflection (i.e., compatible with a path in which the wave velocity is the speed of light in vacuum $c$). For a given high frequency (HF) this quantity can be simply calculated analytically or directly from the ionogram (Davies, 1990). The Breit and Tuve theorem assures that the calculated group time delay $t_{calc}$ of the wave propagating in the effective path at group velocity compatible with the group refractive index is equal to the virtual time delay $t_{virt}$ of the wave propagating along the oblique virtual path at the speed of light $c$, i.e. $t_{calc} = t_{virt}$. Another useful concept derived from the Martyn's theorem relates the oblique virtual reflection height $h'_{ob}$ at a HF frequency compatible with the oblique propagation $f_{ob}$ to the vertical virtual reflection height $h'_{ob}$ at a vertical frequency $f_v$, i.e. $h'_{ob} = h'_v$. In order to test the ray tracing algorithm the two above mentioned propagating times must be considered. In this paper we deal with a ray tracing algorithm derived from 6 differential equations with Hamiltonian formalism in geocentric spherical coordinates. The ray tracing program is written in Matlab for what concerns input and output routines, while the executive computation nucleus derives from the Jones and Stephenson program (Jones and Stephenson, 1974). The 3D ionospheric model used employs both the analytic and the numerical 3D electronic density model elaborated at the Istituto Nazionale di Geofisica e Vulcanologia (INGV). This model derives from the International Reference Ionosfere (IRI) model (Bilitza, 1990) updated with Autoscala profiles obtained from the actual ionograms in the considered region (Scotto, 2009).

**2. Reflector simulation**

In an oblique path (dashed line in figure 1) from the transmitter positioned at the point indicated with A the ray reaches the point B on the ground surface through the ionosphere. We therefore simulate a perfect virtual reflector at an altitude $h'_v$ compatible with the virtual propagation, i.e. at the vertex C of a triangle ACB representing the virtual path.
The relation between the vertical and oblique frequencies is given by the secant law:

$$f_v = f_{ob} \cos\varphi \qquad (1)$$

where $\varphi$ is the incidence angle, $f_{ob}$ is the oblique frequency and $f_v$ is the vertical frequency. The vertical height of reflection $h'_v$ is calculated from the density profile in the middle point between the transmitter and the receiver solving the integration numerically in case of numerical profile as in the following equation:

$$h'_v(f_v) = h_0 + \int_{h_o}^{h_r} n_g(h, f_v) dh \qquad (2)$$

where $n_g$ is the group refractive index, $h_0$ the initial height of the ionosphere and $h_r$ the real height of the vertical path. In our simulation $n_g$ is given by:

$$n_g(h, f_v) = \left[1 - f_N^2(h)/f_v^2\right]^{-1/2} \qquad (3)$$

where $f_N$ is the plasma frequency which has a value about $9\sqrt{N}$ with $N$ indicating the electronic density in *electrons/m³*. It is also easy to obtain $h'_v$ directly from the ionogram that shows pairs of values of virtual height vs. frequency. In this way the height of the reflection point of the virtual path is found both for vertical ($h'_v$) and oblique ($h'_{ob}$) paths as assured by Martyn's theorem if relation (1) is employed. Hence, for each frequency $f_v$ the relative point C can be obtained. In the next paragraph, we will show how to calculate the coordinates of the point R and the two time delays (calculated and virtual).

**3. Calculating the virtual time delay**

The calculated group time delay $t_{calc}$ of the real oblique path of the wave (indicated as dashed curve in figure 1) is, under opportune hypotheses, equal to the virtual time delay $t_{virt}$. The latter can be obtained very easily using a pure geometrical relation if we consider a flat layering ionosphere or approximately spherical surfaces, i.e. electronic density profile dependent only on the altitude, without any horizontal gradient (see figure 2). A monotonic increasing trend of the electronic density profile is also required for the validity of the relation $t_{virt} = t_{real}$, otherwise the valley between E and F layers invalidates the above theorems.

At this stage we have all the conditions to test a ray tracing algorithm employing both analytical and numerical profiles as detailed above. Practically, the ray tracing algorithm employed (Bianchi et al., 2010) integrates numerically the following *6* differential equations (Jones and Stevenson, 1974) in spherical coordinates *r*, *θ*, *φ*:

$$\frac{dr}{d\tau} = \frac{\partial H}{\partial k_r} \tag{4a}$$

$$\frac{d\theta}{d\tau} = \frac{1}{r}\frac{\partial H}{\partial k_\theta} \tag{4b}$$

$$\frac{d\varphi}{d\tau} = \frac{1}{r\sin\theta}\frac{\partial H}{\partial k_\varphi} \tag{4c}$$

$$\frac{dk_r}{d\tau} = -\frac{\partial H}{\partial r} + k_\theta \frac{d\theta}{d\tau} + k_\varphi \sin\theta \frac{d\varphi}{d\tau} \tag{4d}$$

$$\frac{dk_\theta}{d\tau} = \frac{1}{r}(-\frac{\partial H}{\partial \theta} - k_\theta \frac{dr}{d\tau} + k_\varphi r \cos\theta \frac{d\varphi}{d\tau}) \tag{4e}$$

$$\frac{dk_\varphi}{d\tau} = \frac{1}{r\sin\theta}(-\frac{\partial H}{\partial \varphi} - k_\varphi \sin\theta \frac{dr}{d\tau} - k_\varphi r \cos\theta \frac{d\theta}{d\tau}) \tag{4f}$$

where *H* is the Hamiltonian, $k_r$, $k_\theta$, $k_\varphi$ (see figure 3) are the components of wave vector along *r*, *θ*, *φ*, and the independent variable *τ* must be a monotonic increasing quantity. The Hamiltonian *H* is a constant during the ray propagation and in our algorithm the following relation has been chosen:

$$H(r,\theta,\varphi,k_r,k_\theta,k_\varphi) = \frac{1}{2}\text{Re}\left[\frac{c^2}{\omega^2}\left(k_r^{\,2} + k_\theta^{\,2} + k_\varphi^{\,2}\right) - n^2\right] \tag{5}$$

where *n* is the phase refractive index of a simple medium without magnetic field and without collision, derived from relation (3). The best choice for *τ* is *τ=c·t* that in discrete steps is *Δl=Δt·c*. This last relation furnishes directly the calculated delay $t_{calc}$ using:

$$t_{calc} = \sum \Delta t \tag{6}$$

that will be compared with the virtual delay $t_{virt}$.

The ray tracing algorithm ends the integration when the ray reaches the ground after reflection. At this point we have the coordinates of the point R, i.e., $r_R$, $\theta_R$, $\varphi_R$. So the virtual time delay $t_{virt}$ can be immediately obtained as: $t_{virt} = ACB/c$. We tested the algorithm for two different paths up to thousands of *km* using two different elevation angles and the two different ionospheric electron density models (numerical and analytic). If we choose a relative short path so that the spherical geometry in figure 1 can be approximated to the flat Cartesian geometry of figure 2, the above theorems are better approximated and the errors can be only ascribed to the discrete step integration of the employed algorithm.

**4. Calculation of time delay differences by analytic and 3D numerical models.**

We simulate reflectors placed at different heights compatible with the various vertical frequencies $f_v$ considered. We find that the ray tracing algorithm fits nearly perfectly the theory established by the above mentioned theorems in case of an analytic, monotonically increasing, electronic density profile such as the Chapman one, but also in case of parabolic or linear profiles (Budden, 1985). This means that the relative errors $\Delta t_{error}$ between the calculated $t_{calc}$ and the virtual $t_{virt}$ time delays are almost negligible. Figure 4 shows $\Delta t_{error}$ corresponding to an analytic Chapman model for the elevation angle of *30* degrees. We notice that $\Delta t_{error}$ does not exceed *0.2%* when the frequencies are relatively low. We also point out that the error increases when the ray penetrates deep in the ionospheric plasma since numerical integration meets the discontinuity in the refractive index next to the critical frequency. For the numerical model shown in figure 5 the simulations are performed for three different elevation angles, *18*, *30* and *45* degrees, as shown in figure 6, 7 and 8, respectively. The red line indicates virtual time delay, blue line indicates the calculated time delay, while the black line indicates the percentage error between the virtual and calculated delays. We found the same general trend as in the analytic model except for the more irregular behavior due to the discrete values in the electronic density profile with a step of *0.5 km*. Approaching the frequencies of penetration relative errors increase because of the finite step in discrete integration process. For a numerical electronic density profile we used the relation (2) to obtain the vertical virtual height $h'_v$ at the correspondent frequency $f_v$, where the latter is a parameter during the integration. Relative errors are greater with respect to the analytic electronic density profiles because of an additional discrete step integration taking place in the whole process. Nevertheless the relative errors are always less than *1%* if the maximum electronic density is not reached.

**5. Conclusions**

We employed a simple method to test a ray tracing algorithm. The relations above described were exploited to obtain a quantity replacing the measured group time delay $t_{real}$ that is a value obtainable from an actual measurement. The last quantity can be compared with the virtual time delay $t_{virt}$ to check the performance of the algorithm. A perfect reflector placed at an altitude equal to the virtual height of reflection was simulated to calculate the virtual time delay that is assumed as a real time delay. For monotonic electronic density profiles, we found a very small relative error between the calculated and the virtual delays for both analytic and numerical 3D electronic density models. In order to start our analysis, the percentage relative errors between the calculated and the virtual time delays were checked using an analytic model defined by a linear electronic density profile. Very small relative errors were always of magnitude order less than *0.2%*. Such errors are mainly due to the discrete step of the numerical integration of the differential equations of the ray tracing algorithm and to numerical electronic density profiles. A further analysis has been performed checking the resulting virtual height of the simulated reflector and the virtual height calculated by means of the relation (2). The small errors between the two values were always less than *1%* when the interested ionospheric region does not reach the maximum electronic density.

**Figure captions**

Figure 1. Reflector positioned at the height $h'_v$. Once the elevation angle of radiation $\beta$ and the points A and B are known from the output of the ray tracing, all the geometrical parameters are calculated using the following relations: $i = R_T[\sin(\theta/2)/\sin\varphi]$, $\varphi = \pi/2 - (\theta/2+\beta)$, $h'_{ob} = i\cos\varphi - d$, $d = R_T[1-\cos(\theta/2)]$, where $R_T$ is Earth's radius while the other parameters are evident in the figure.

Figure 2. Flat stratified ionospheric medium described by an analytic linear model.

Figure 3. Projection of wave vector **k** along $r$, $\theta$, $\varphi$ (versor $i_r$, $i_\theta$, $i_\varphi$) in spherical coordinates.

Figure 4. Time group delays and percentage relative error simulating a α-Chapman's electronic density profile. The dashed- red curve is the virtual time delay $t_{virt}$ and the blue curve the calculated time dealy $t_{calc}$ in ms, while the black curve indicates the percentage relative errors at the various employed frequencies in *MHz*. The elevation angle is *30* degrees.

Figure 5. Numerical electronic density profile defined by parameters such as critical frequency and height of maximum electronic density.

Figure 6. Time group delays and percentage relative error simulating a numerical electronic density profile (refer to caption of figure 5) with elevation angle of *18* degrees.

Figure 7. As in figure 6 with elevation angle of *30* degrees.

Figure 8. As in figure 6 with elevation angle of *45* degrees.

**Figure 1**

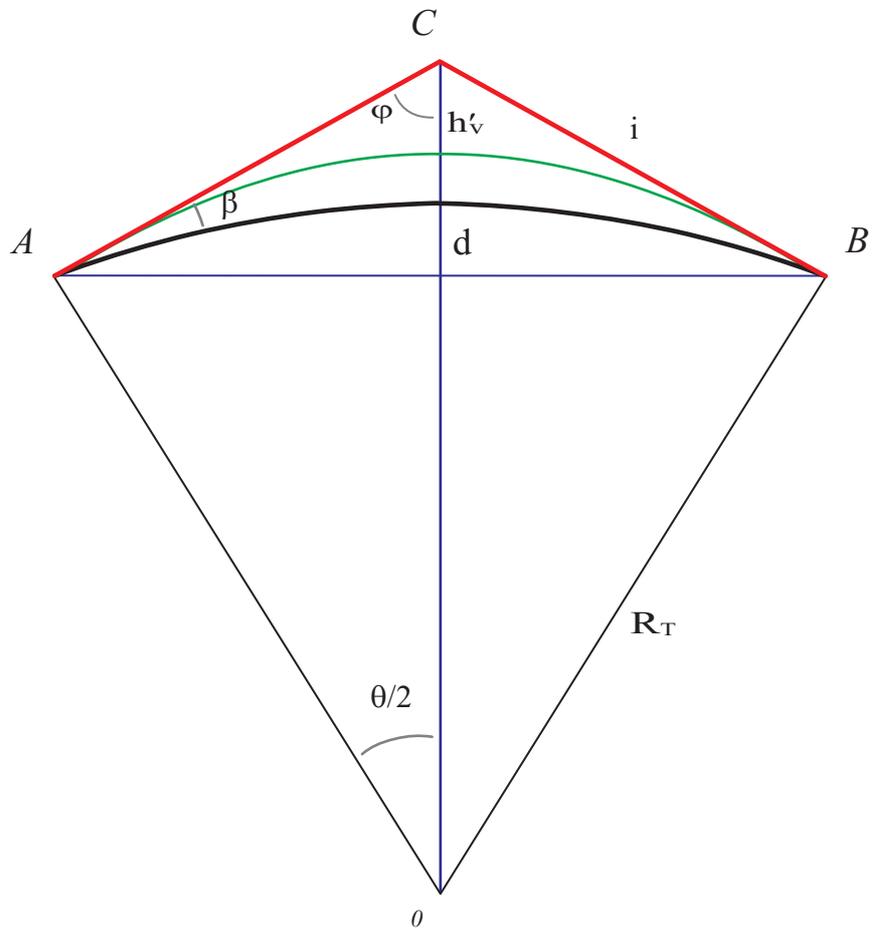

**Figure 2**

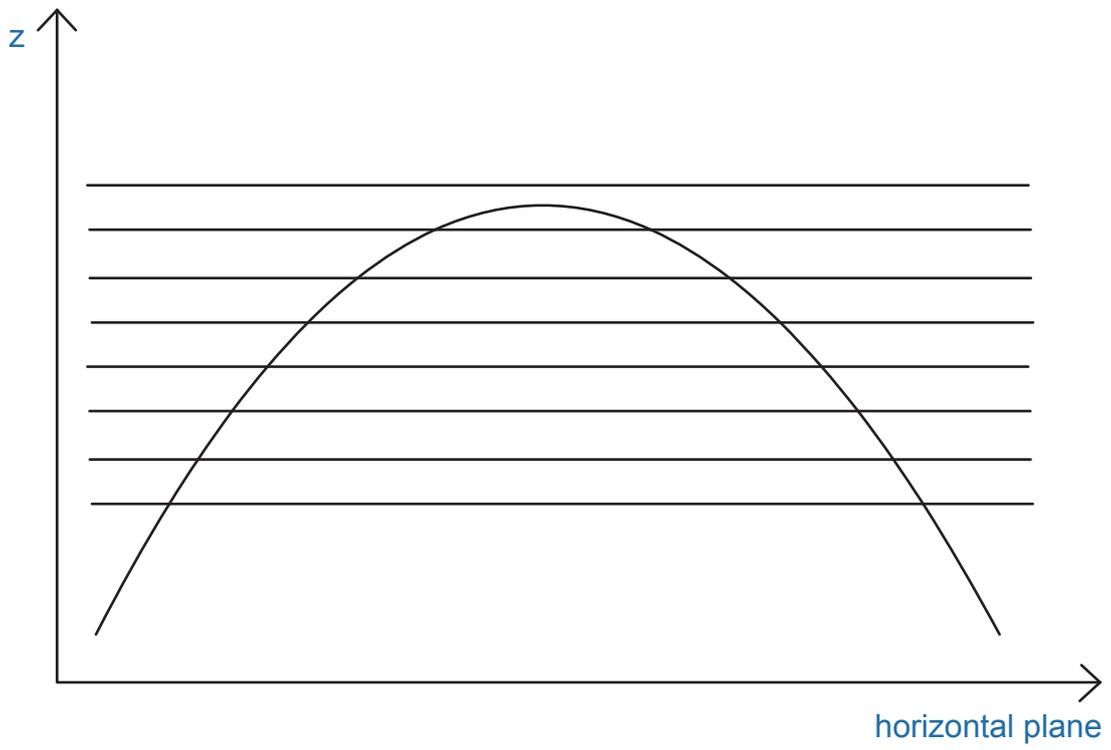

**Figure 3**

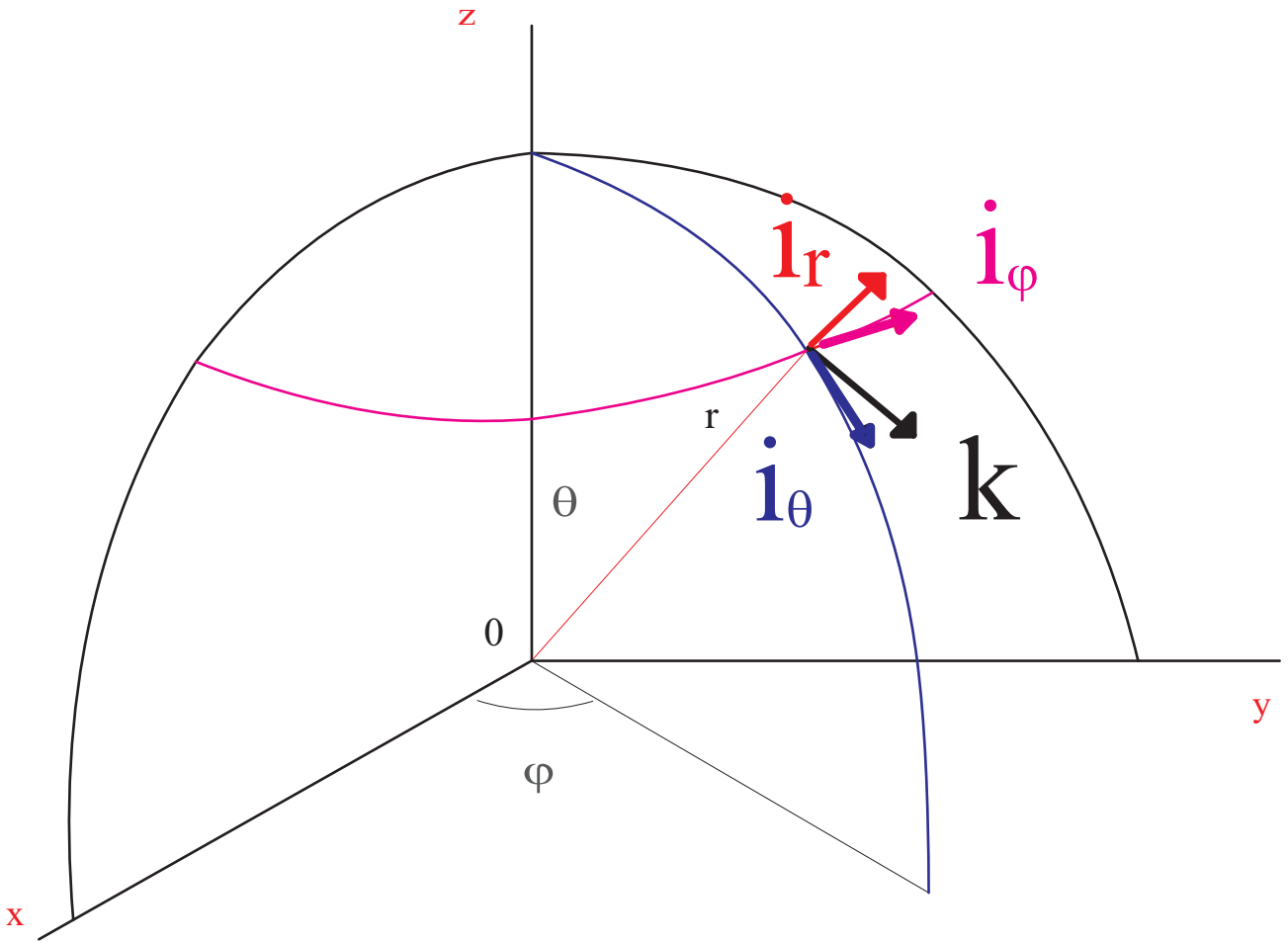

**Figure 4**

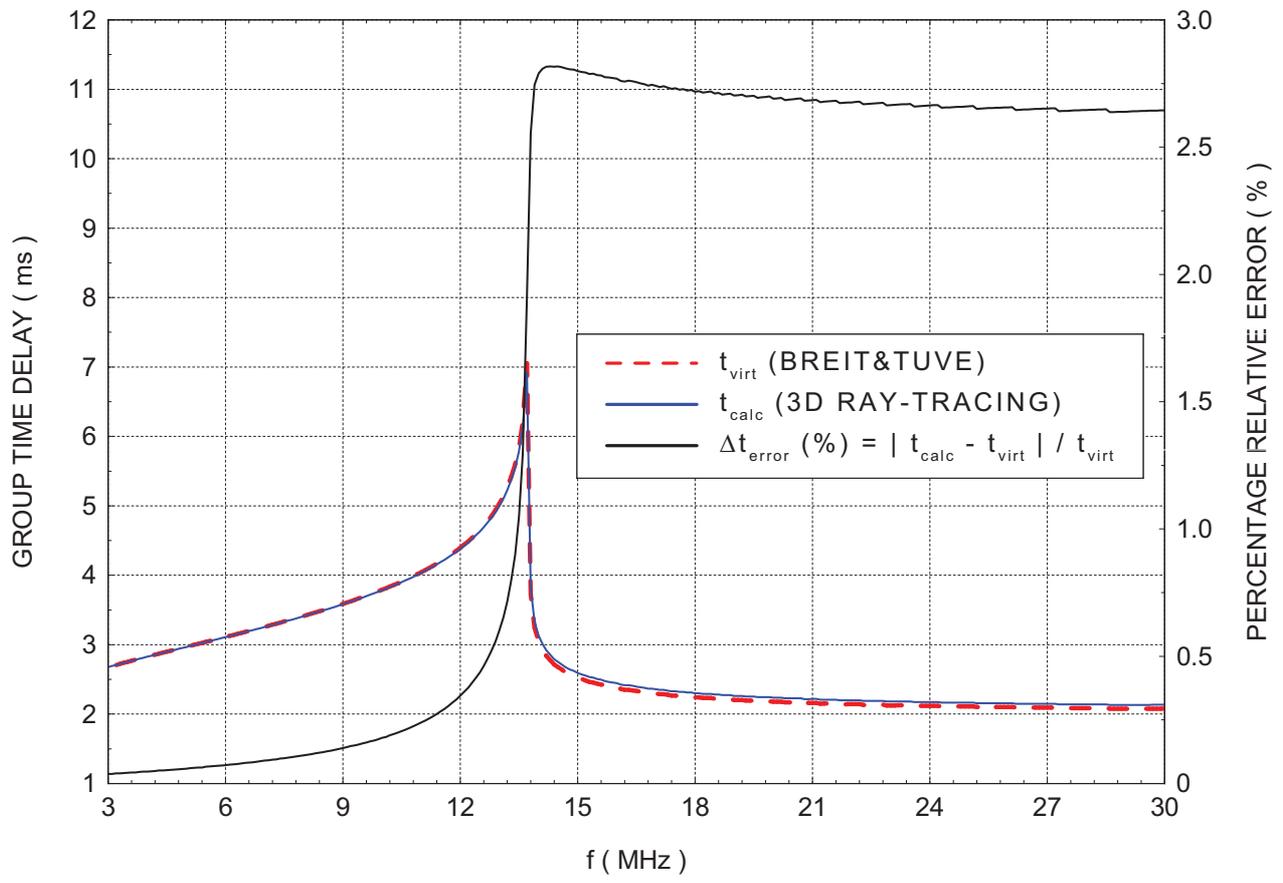

**Figure 5**

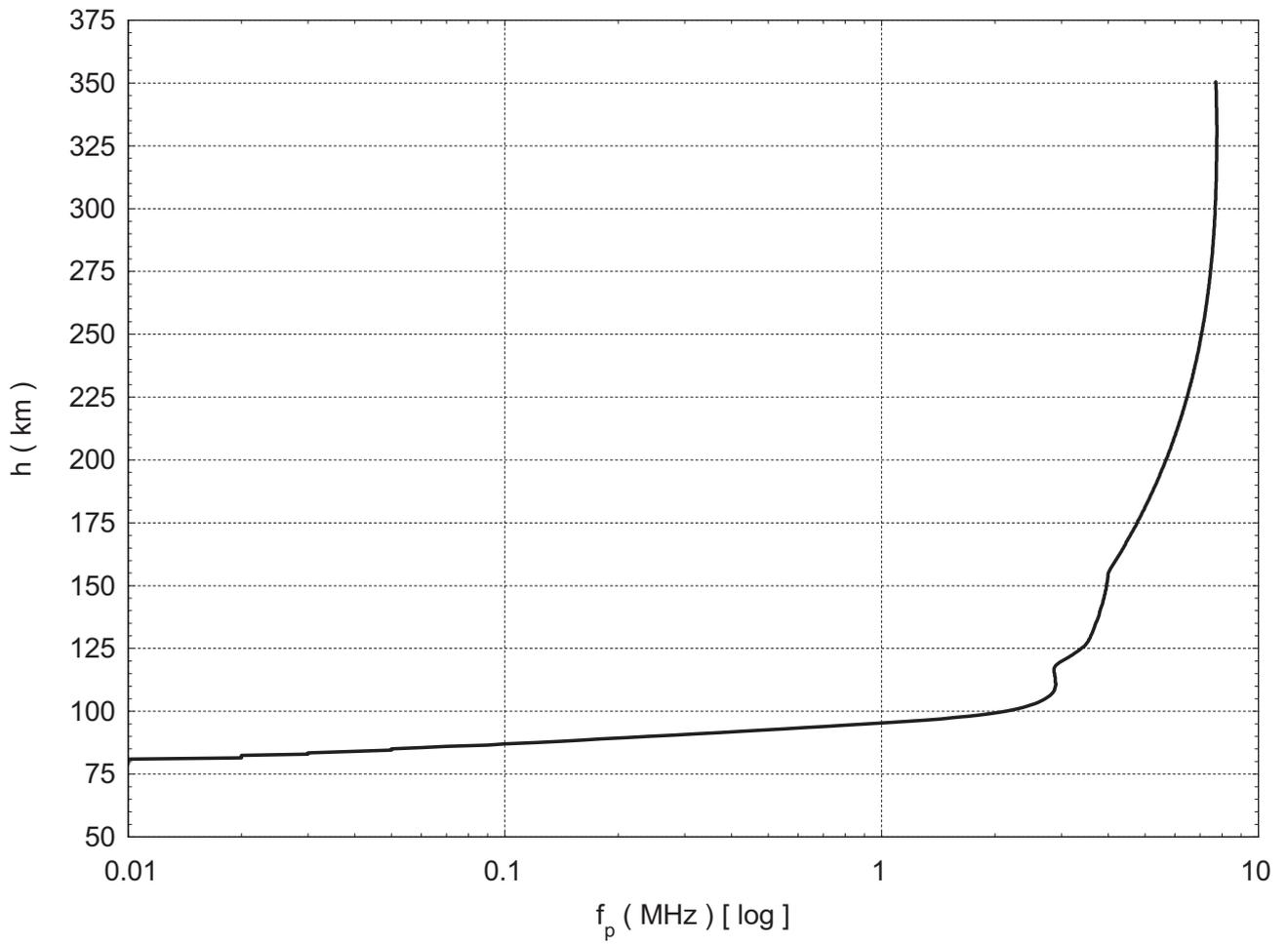

**Figure 6**

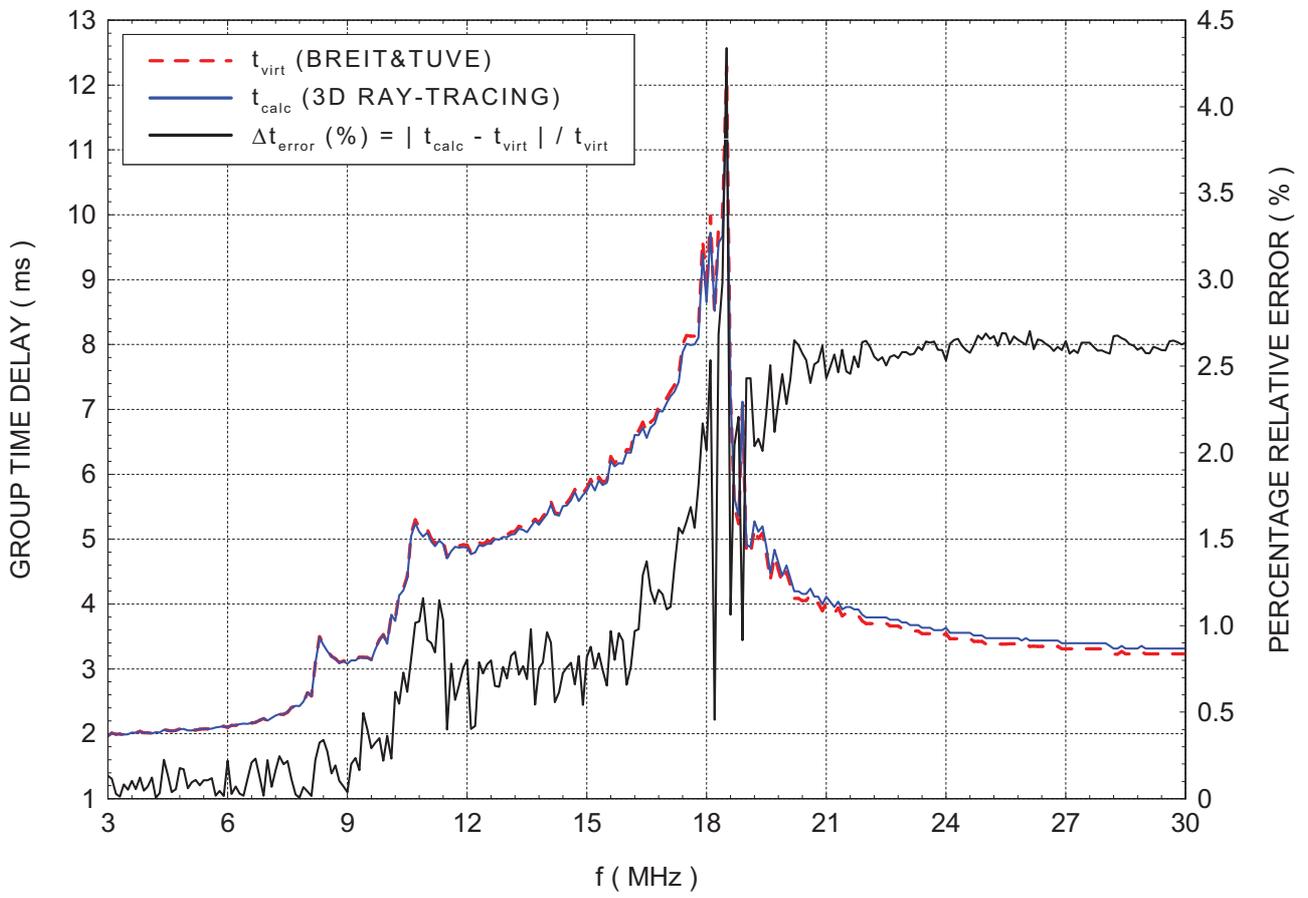

**Figure 7**

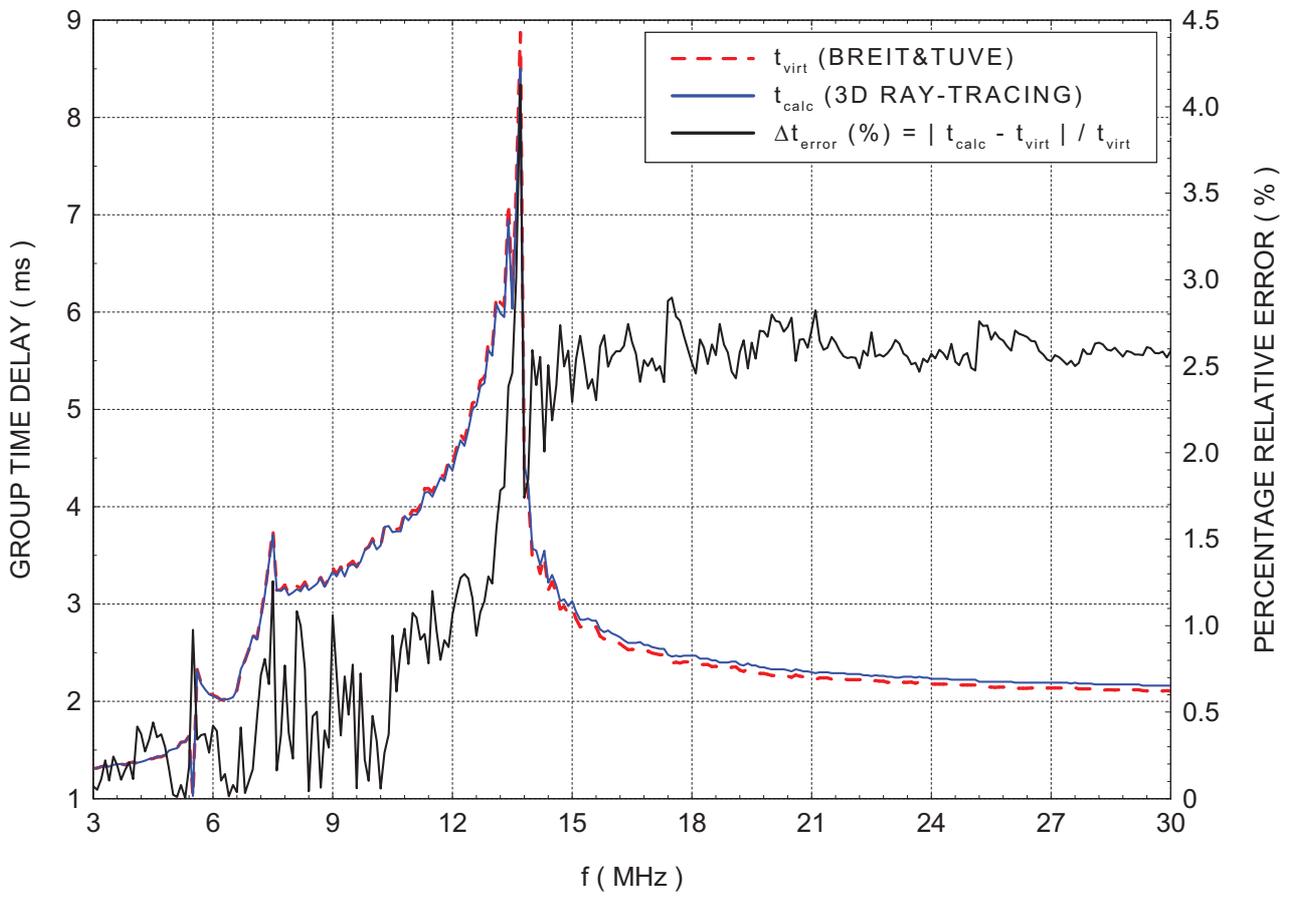

**Figure 8**

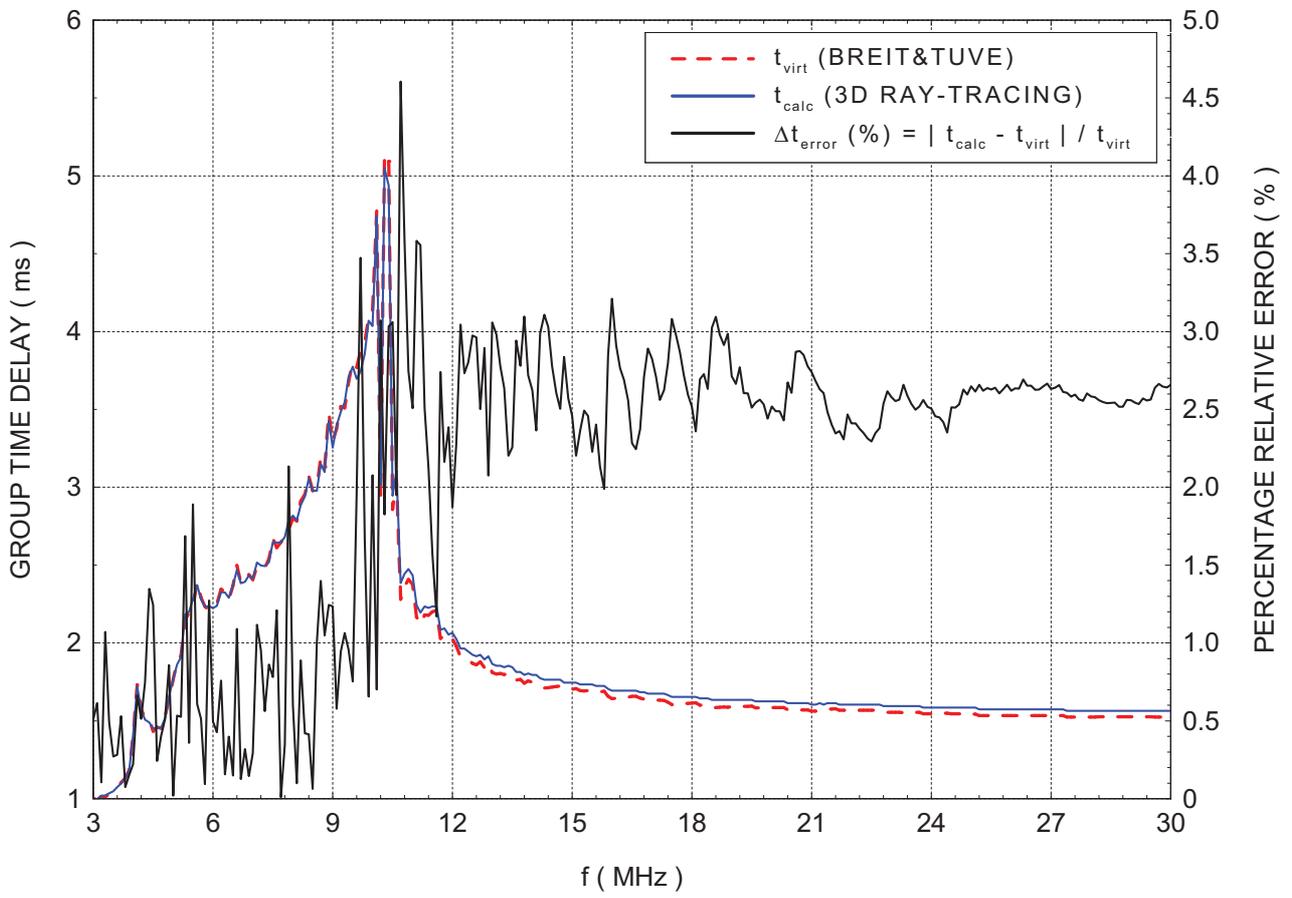